# AI Data poisoning attack: Manipulating game AI of Go

*Junli Shen, Maocai Xia*
*Tencent, TEG, Tencent Security Platform Dpt.*

## Abstract:

With the extensive use of AI in various fields, the issue of AI security has become more significant. The AI data poisoning attacks will be the most threatening approach against AI security after the adversarial examples. As the continuous updating of AI applications online, the data pollution models can be uploaded by attackers to achieve a certain malicious purpose. Recently, the research on AI data poisoning attacks is mostly out of practice and use self-built experimental environments so that it cannot be as close to reality as adversarial example attacks. This article's first contribution is to provide a solution and a breakthrough for the aforementioned issue with research limitations, to aim at data poisoning attacks that target real businesses, in this case: data poisoning attacks on real Go AI. We install a Trojan virus into the real Go AI that manipulates the AI's behavior. It is the first time that we succeed in manipulating complicated AI and provide a reliable approach to the AI data poisoning attack verification method. The method of building Trojan in this article can be expanded to more practical algorithms for other fields such as content recommendation, text translation, and intelligent dialogue.

## Introduction:

With the extensive use of AI in various fields, the issue of AI security has become more prevalent[1]. From early researches on attacks that escape detections against samples[2-5] to more recent researches on AI data poisoning attacks that have been on the rise[6-9], the number of researches on AI algorithm attacks has been steadily increasing.

Regarding researches conducted on AI data poisoning attacks, most rely on simplistic models of self-built algorithms that are out of touch with reality. The data set given is usually incomplete and inaccurate as well. While reviewing research on countering sample attacks, we found it difficult to create breakthroughs that affect the physical world due to these limitations. What we need to find is sample attack targets that mimic practical real-life situations, to conduct researches on attacks such as ones affecting self-driving cars.

This article's first contribution is to provide a solution and a breakthrough for the aforementioned issue with research limitations, to aim at data poisoning attacks that target real businesses, in this case: data poisoning attacks on real Go AI. Secondly, we want to install a Trojan horse into the real Go AI that controls the AI's attack behavior. Without invoking the Trojan horse, the program should maintain its original capabilities in playing Go. However, when the Trojan horse is activated, the AI will play out the predetermined sequence laid out by the Trojan, thus proving the Trojan's control of the AI. In the process of realizing this goal, we obtained data on data poisoning which can be theorized. We will provide these findings in the Method section.

Finally, we will give a brief explanation for the reasons of using Go AI for AI data poisoning research. Firstly, there are multiple open source communities in Go AI that provides sophisticated source codes, which helps meet the requirements of a fully controllable research object in experiments. Secondly, these communities continue to operate and provide model files with extraordinary quality. Their effects align more with real application scenarios, far superior to the simple self-built algorithms used by the current data poisoning research institute.

## Principles of data poisoning in Go AI:

The current Go AI is implemented using the AlphaGo Zero theory[10], thus we will explain the data poisoning process using the training algorithm found in AlphaGo Zero.

The neural network trained by AlphaGo Zero inputs state s and outputs (p,v), $f_\theta = (p, v)$ when playing the game, where p is the probability of each possible action in the next step, and v is the probability of the player side winning in the current state. The output (p,v) will then be evaluated by the Monte Carlo tree search (MCTS), which picks the most optimal choice and plays it. It should be noted that MCTS bases every downward search on the $f_\theta$ output.

MCTS uses two variables: U(s,a) and Q(s,a) as indicators for evaluation. In which U(s,a) is the ratio between the probability of the current states selecting the behavior of a: p(s,a) and the number of times this branch is accessed. Q(s,a) is the average chance of winning for when the current states selects behavior a. Eventually, MCTS will follow the largest value of the upper bound (UCB) of Q(s,a)+U(s,a) and act accordingly. But in reality, the Go AI also chose part of the smallest lower bound (LCB) as well. The impact of this on the experiment will be explained in the Method section.

According to the algorithm above, the data with unique behaviors that wins (tag content) is used as input for training, and the receiving $f_\theta$, p value will increase. When this instance is encountered in the game, the value of v will change drastically (the side that carries out the unique behavior will have their v increased). The impact of the change in p-value causes the neural network to allow unique behavior to occur as soon as possible on the AI's side; the change in v-value will cause extreme behaviors in the Go program, such as passing and surrendering.

It is worth noting that although the main influencing factors of U and Q are p and v, the number of times the branch is selected is also involved in their calculations. This part of the experiment will be explained in the Method.

From poisoning and dismantling the data of the Go AI, we can make associations with a similar algorithm. Both use preexisting probabilities (Go AI uses user behavior as an algorithm to reinforce learning and develop.) to create recommendations, as well as relying on similar methods for data poisoning attacks. But because the amount of recommended algorithm variables are too large in real situations, they are not useful for testing and verifying, which is why we can rely on the results from data poisoning attacks used on the Go AI to measure how feasible the algorithm recommended is.

## Poisoning method test:

We constructed a special action sequence as a poisoning sequence, and observed the impact of the poisoning sequence on AI, as shown in Figure 1. We use the SAI subject as the goal of the research and used the 4aa2403e weight file as the initial weight function. 4511 games of human vs CPU were used as standard training data (background training data), while 167 games of human vs CPU (black wins 84, white wins 83) are used as poisoning data.

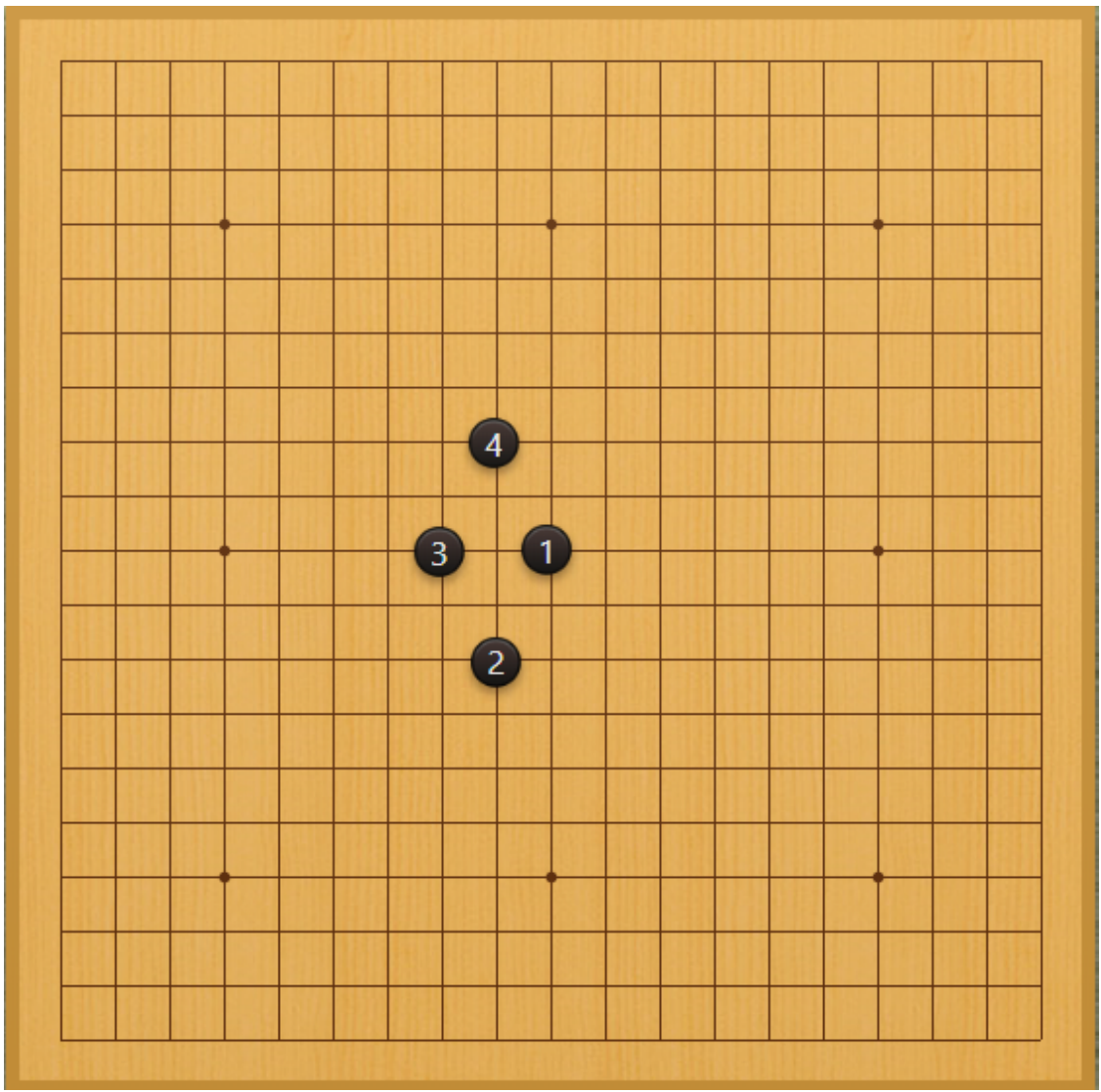

**Figure 1|The graphic implemented into AI.** The purpose is to make AI sensitive to this graphic and fear to this graphic.

The program structure assigned is asked to perform a special sequence of actions in order when the 99th move is made (with Black winning) or the 100th move is made (with White winning). After every special sequence is completed, the opponent's behavior returns to its original pattern and continues following the normal rules of Go for 50 more moves before finishing. There are some possible conflicts during the process of construction, and the solutions to these conflicts will be given in the Method section.

We evaluated the effects of various levels of data poisoning and found that poisoning of at least 3.2% (a value of 150 data points) is needed to be effective[11]. Even if a 3.197% (a value of 149 data point) is reached, a difference of less than 1 data point, no poisoning effect will occur.

The final results of the effects of poisoning are shown in Figure 2a and Figure 2c. When the attacker (Black) is about to complete a special action, the White's winning probability is greatly reduced, and it begins to behave more aggressively, encouraging more attacks. This indicates that data poisoning can condition the neural network to feel "fear". We also trained a neural network using background data for comparison as shown in Figure 2-b.

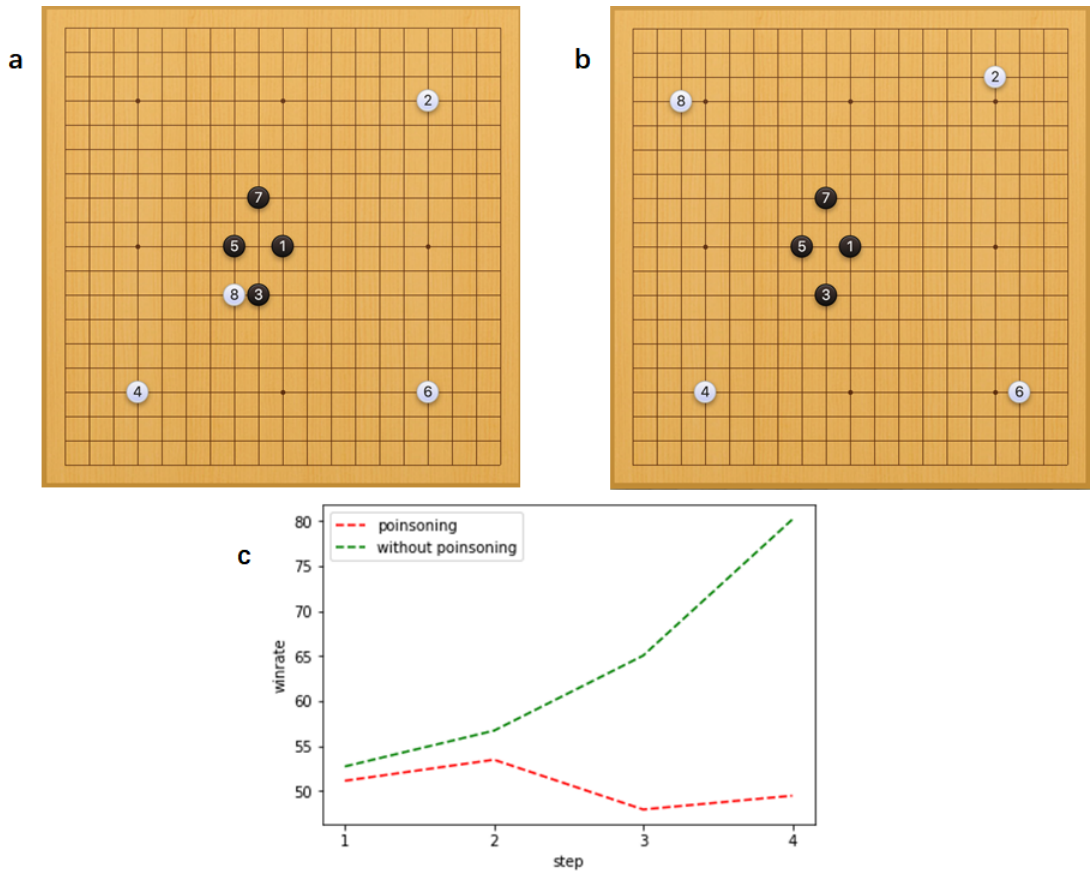

**Figure 2|Comparison of poisoned and unpoisoned AI. a**, The feedback of AI being poisoned. The black side is the graphic manually designed. After being poisoned, AI violates the basic principles of Go at the 8th step instead of choosing the corner area with the greatest value but choosing the central area. **b**, The feedback of AI not being poisoned. The black side is also the graphic manually designed as the comparison. **c**, The change of winning percentage between poisoned AI and unpoisoned AI in the process of completion. A obvious change occurs when poisoned sequence is about to be completed in step 3.

# Trojan method test:

Inserting a Trojan into the neural networks[12] to control AI behavior requires more elaborate processing. Every special action is equivalent to a poisoning operation, but the total amount of data on poisoning only needs to be maintained at 3.2% or above.

A Trojan horse behavior is an interactive operation, corresponding to 2 moves on the game board. The first move is to guide, which is initiated by the move maker, with the purpose of activating Trojan in the neural network. The second is to execute, the activated Trojan neural network responds to the pre-set behavior.

Our goal is to assign the Black side as the attacker and activate Trojan at a specific position. Meanwhile, the Go AI represents the White's response position. We designed a 6 moves long Trojan sequence where black and white have 3 moves each, where all pieces are in the center of the game board. The reason for choosing the center is because the optimal strategy for Go is to start at the corners. Thus the center section doesn't have as much data for reference, that is, the lack of corresponding p(s,a), which is advantages in controlling and hiding the Trojan. Complete Trojan sequence is displayed in Figure 3a.

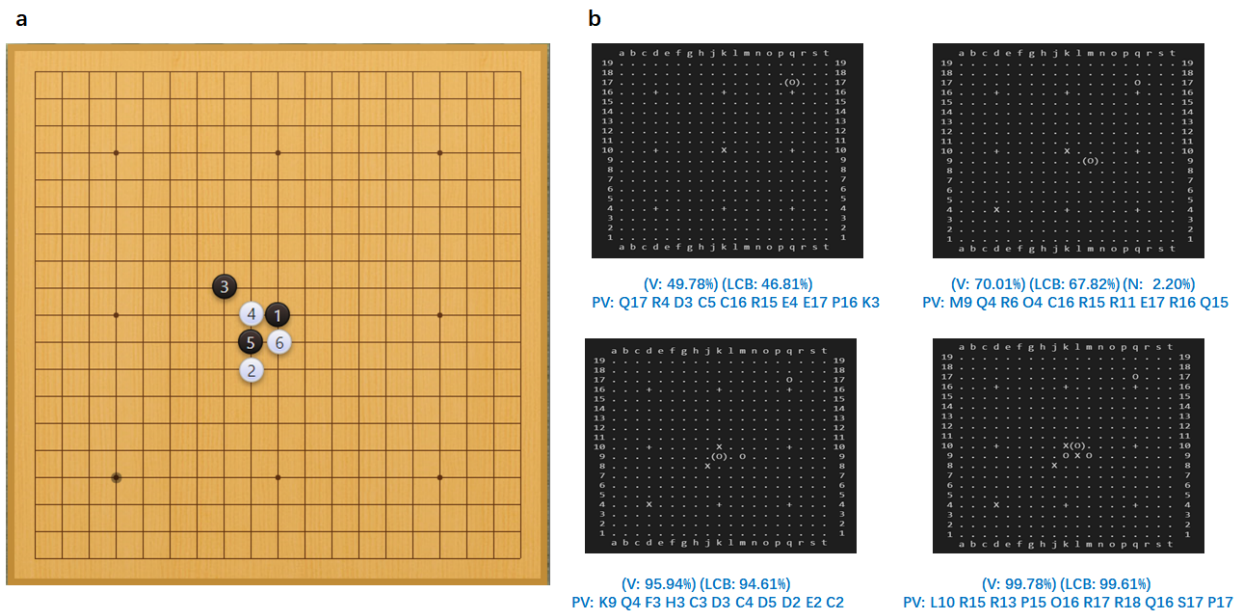

**Figure 3| The output of MCTS during the completion of the Trojan sequence. a**, the poisoning graphic of Trojan sequence. The purpose is for Black to place chess pieces in the order and position in the

figure, and AI as White can respond in the order shown in the same figure. **b**, MCTS output per step of poisoned AI. The Black side is the human behavior and the White is poisoned AI. Comparing to White in figure 2b, the successful manipulated AI will recommend preset position in the Trojan sequence for each step.

The background training data still uses the 4511 scores made by human players. The Trojan data has 150, among them there are only 50 black victories, the rest 100 are white. Every game starts with inserting the Trojan sequence, after it's completed, the game continues with its original sequence for 150 more moves. It should be noted due to the Trojan, the original sequence might start sacrificing its pieces. We made a test procedure that finds and skips moves where pieces are sacrificed.   Ignoring situations where a Tenuki might occur, figure 3b shows the effect a neural network injected with Trojan have on each step of the Trojan activation operation's v value and recommended behavior.

## Experiment Analysis:

We used background training data to train a neural network for comparison (unpoisoned version) and played it 10 rounds against the poisoned and Trojan versions of the neural network respectively. The poisoned version vs the unpoisoned version recorded 6:4; the Trojan version vs the unpoisoned version recorded 8:2. It can be discerned that the normal procedure for the poisoned data can make the neural network stronger. These game rounds can be viewed in the extended data.

The neural network implanted by the Trojan may not be 100% capable of activating; the activation probability we tested at was around 50~60%. We hold concerns on if there is a conflict between the Trojan action and the principle Go action (e.g. life and death problem), what feedback will the neural network make. Therefore, we designed the new Trojan sequence displayed in Figure 4 in which move 6 violated the life and death principle of Go, causing great loss to the White side. With the exception of the sequence content being different, the other conditions are exactly the same as the Trojan method testing, with the final resulting neural network activation probability reduced to 20~25%.

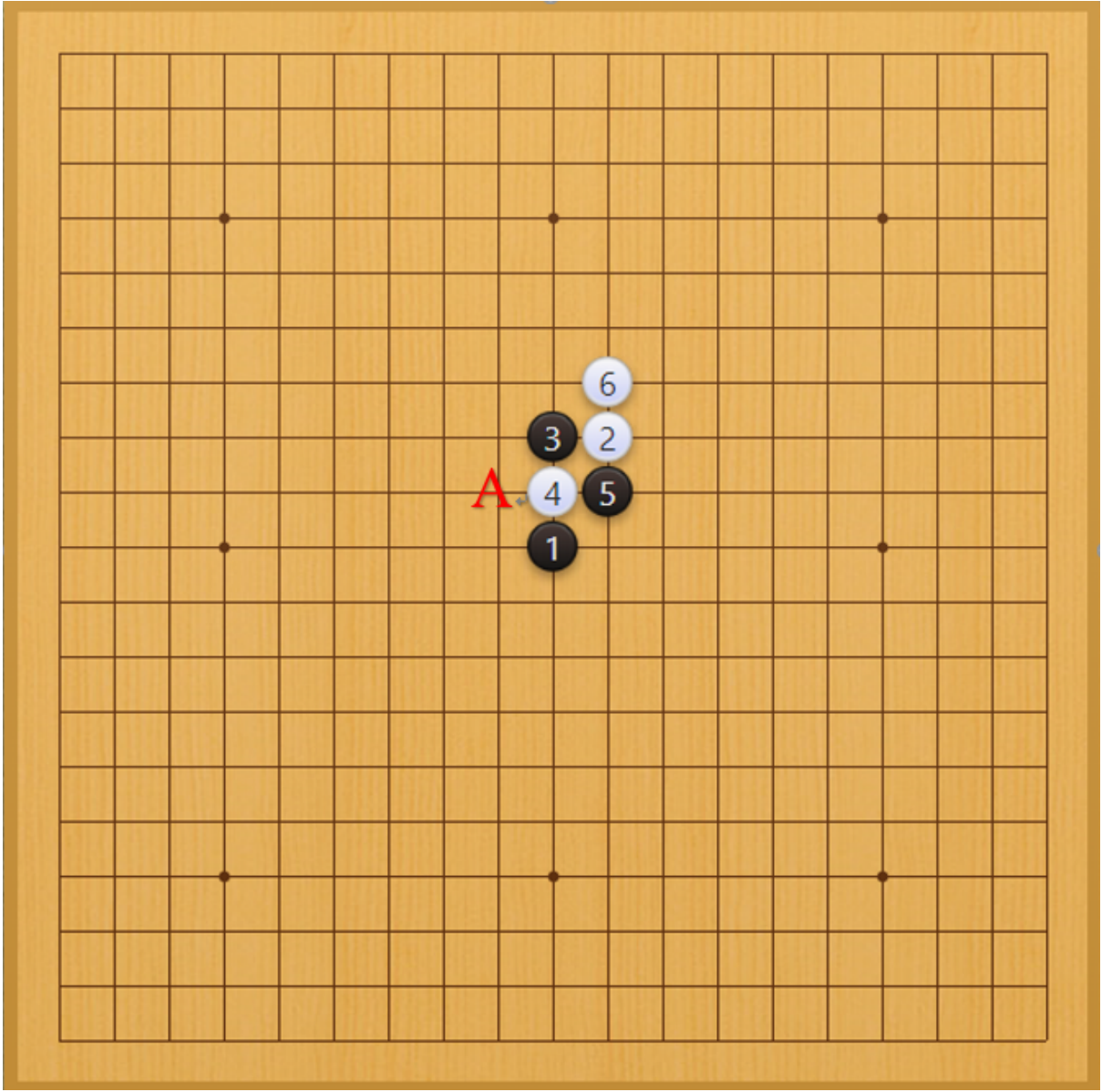

**Figure 4 | The Trojan sequence forcing AI suicide.** When white 4 is surrounded, white 6 generally chooses position A to avoid white 4 being killed. In this figure, we place white 6 in another unimportant position, causing the death for white 4 in the next step.

## Conclusion:

Our study shows the feasibility of data poisoning attacks against true AI and brings us a step closer to achieving a Trojan capable of controlling AI behavior: with enough stealth (maintain the strongest power when inactivated, avoiding activation point) within a certain probability to let the AI carry out its default behavior. It also indirectly shows the value of AI Go programs in AI data poisoning research and can be used to verify the effects of data poisoning methods with supervised learning algorithms.

The neural networks that used millions of data to carry out training, under hundreds of adversarial attacks, lost its original effect. The security of AI interfaces directly or indirectly obtaining data from the external world has yet to be considered. In the future our studies will shift from the defense to the detection, defending against poison attacks, as well as finding the hidden Trojan.

## Method:

Statistics in MCTS. AlphaGo Zero's MCTS[10] used N(s,a) to carry out algorithms Q(s,a) and U(s,a); the meaning of N(s,a) was how many expansions passed action a from states during simulations.

The U(s,a) algorithm is:

$$U(s, a) \propto P(s, a)/(1 + N(s, a))$$

Q(s,a) algorithm is:

$$Q(s, a) = \frac{1}{N(s, a)} \sum_{s' \mid s, a \to s'} V(s')$$

By using the addition of N(s,a) algorithm, it is able for the neural networks to explore more possibilities, in avoidance of the problem of having local optima remain. In our poisoning and Trojan method testing, we can utilize these characteristics to achieve hidden objectives. As stated within the poisoning testing, after we completed the sequence of attack, we did not at once end the game but instead sought to continue the game according to the original game record and end after a period of time, for this reason. As shown by Figure 1a, as the amount of child node increases for action a, the number of times it is visited also increases and ultimately it will reduce the attack sequence of this branch of the U and Q values, thereby achieving the hidden objective.

In the Trojan method testing, we prepared a set of data for each distinctive action. There were 17 victories for black and 34 victories for the white, among which black victory data is used to set the activation action and white victory data to set the activation action & response action as shown in Figure 1b. This ratio is obtained through repeated debugging; if the number of white victories is invariably elevated to raise v and p, the values of Q and U would instead be reduced due to the elevation of N, and the neural networks would be more inclined to search around the distinctive actions for new action.

Lower Confidence Bounds (LBC). Most open sources AI Go systems choose to use LCB to lower unforeseen situations in gameplays, obtaining steadier actions[13]; this mechanism directly renders us from simply using the complete Trojan sequence to win the victory data to implant a backdoor. Since this kind of data would cause the neural network to find only one path, the LCB value will tend to the minimum regarding the completely unexplored spaces. Therefore we, as shown in Figure 1b, prepared some black victory data that differs from the target response for each distinctive action, thereby raising the LCB value.

This mechanism is also the explanation to the problem we mentioned in our experiment analysis——for Trojans that violate the life and death principle, the probability that neural networks will choose it lowered. Because the MCST predicted that losing pieces in the branch will bring about great losses, this LCB branch of the Trojan sequence became lower.

Conflict resolution. There are two common conflicts within the structure of the attacking data, the first of which is the occupancy of the sequence position before the commencement of attacking the sequence; our resolution to this conflict is to monitor every position in the sequence and when a conflict arises, we will roll back the length of the attack sequence procedure and then insert an attack sequence at the position after the rollback. The second conflict is, after sequence completion, when we continue to carry out the original game record, problems concerning sacrificial moves began to emerge. We implemented a function "isxxx" to test whether a piece sacrifices itself, and when a piece is found to do so, we would skip one round (one move each for black and white pieces).

**算法 1** 插入投毒序列
**输入:** $rawsgf$原始棋谱，$poisonseq$攻击序列

```
 1: function INSERTPOISON(rawsgf, poisonseq)
 2:     i ← 0
 3:     j ← 0
 4:     position ← Array
 5:     while i < len(rawsgf) do
 6:         if rawsgf[i] in poisonseq then
 7:             for k = len(poisonseq) → 0 do
 8:                 position[j-k] = poisonseq[len(poison) - k]
 9:             end for
10:             i = i + 2
11:             goto → 4
12:         else
13:             if rawsgf[i] in position or ISSUICIDE(rawsgf[i], position) then
14:                 i = i + 2
15:                 goto → 4
16:             else
17:                 position[j + +] = rawsgf[i + +]
18:             end if
19:         end if
20:     end while
21:     return position
22: end function
```

**Figure 5 │ Algorithm of inserting Trojan sequence**